\documentclass[twoside,leqno,twocolumn]{article}

\usepackage[letterpaper]{geometry}
\usepackage{subcaption}    
\usepackage{ltexpprt}
\usepackage{graphicx}
\usepackage{url}
\usepackage{xcolor}
\usepackage{enumitem}

\newcommand{\portf}{\textsc{PortFiler}}

\newcommand{\neta}{\textsc{Net-A}}
\newcommand{\netb}{\textsc{Net-B}}

\begin{document}

\title{\Large Collaborative Information Sharing for ML-Based Threat Detection}

\author
{Talha Ongun\thanks{Northeastern University, Boston, MA, USA}
\and Simona	Boboila\footnotemark[1]
\and Alina Oprea\footnotemark[1]
\and Tina Eliassi-Rad\footnotemark[1]
\and Alastair Nottingham\thanks{University of Virginia, Charlottesville, VA, USA}
\and Jason Hiser\footnotemark[2]
\and Jack Davidson\footnotemark[2]
}
\date{}

\maketitle








\begin{abstract} 

Recently, coordinated attack campaigns started to become more widespread on the Internet. In May 2017, WannaCry infected more than 300,000 machines in 150 countries in a few days and had a large impact on critical infrastructure. Existing threat sharing platforms cannot easily adapt to emerging attack patterns. At the same time, enterprises started to adopt machine learning-based threat detection tools in their local networks. In this paper, we pose the question: \emph{What information can defenders share across multiple networks to help machine learning-based threat detection adapt to new coordinated attacks?} We propose three information sharing methods across two networks, and show how the shared information can be used in a machine-learning network-traffic model to significantly improve its ability of detecting evasive self-propagating malware.

\end{abstract} 

\section{Introduction}

With the increased connectivity of devices on the Internet, attackers have an opportunity to launch coordinated global attacks targeting multiple networks. Self-propagating malware (SPM) such as Mirai~\cite{antonakakis2017understanding} and WannaCry~\cite{wannacry} infected hundred of thousands of machines and caused significant damage on a global scale.   More recently, a widespread campaign in the supply chain of the SolarWinds remote monitoring software infiltrated thousands of organizations around the world~\cite{solarwinds}.


Public threat intelligence platforms such as Malware Information Sharing Platform (MISP)~\cite{misp} provide a database of malware signatures and indicators of compromise (IoCs) to enable network owners to deploy solutions to protect against known threats. These platforms can be used to detect known threats, but they fail to adapt quickly to new attack campaigns. Moreover, adversaries can easily create many malware variants in an attempt to evade static signature detection. Recently, machine learning-based threat detection techniques have been proposed (e.g.,~\cite{beehive,buczak2015survey, xin2018machine,made}) and they have started to be widely adopted in the industry~\cite{symantec2020Jun, microsoft2020Feb}.  Most of these ML systems train models locally on the enterprise network, and attempt to detect attacks using the security logs of a single enterprise.

In this paper, we pose the question:  \emph{What information can defenders share across  networks to help machine learning-based threat detection adapt to new coordinated attacks?} Our hypothesis is that ML-based tools can significantly increase their detection ability by leveraging attack information shared across multiple defenders. We demonstrate the advantages of information sharing via a case study of detecting SPM attacks using the \portf\ network traffic system~\cite{portfiler}. We propose and analyze three methods for information sharing, including (1) sharing an entire ML model; (2) sharing weights for an ensemble model; (3) sharing weights for an ensemble model and feature distribution information. All of these methods share aggregated information and protect the privacy of security logs, an important consideration in sharing threat information. Our analysis shows that a locally-trained ML ensemble can detect more evasive malware with $0.91$ precision at  $0.86 $ recall on average using shared threat information and feature distribution information. We conclude by discussing the challenges on designing and deploying global defensive ML models to counteract coordinated attacks.

\section{Methodology}


\begin{figure*}[t]
	\centering
	\includegraphics[width=0.83\linewidth]{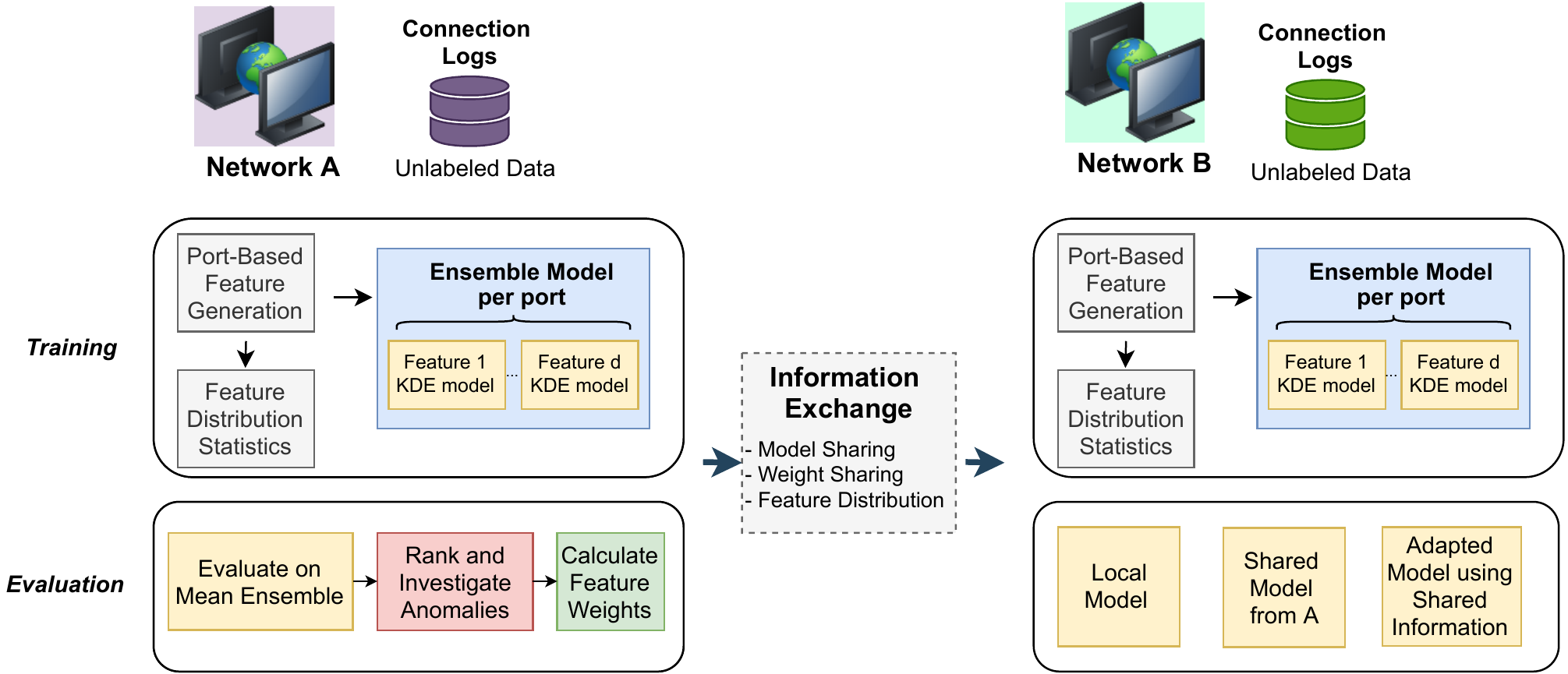}
	\caption{Overview of the information sharing platform where \neta\ shares either its local ML model, feature weights for the ensemble, or feature distribution to improve the local ML  model of \netb. }
	\label{fig:global}
	
	\vspace{-4mm}
\end{figure*}


We provide an overview of the \portf\ system designed to proactively detect SPM attacks in an enterprise network~\cite{portfiler}. We discuss our machine learning (ML) methodology, and three information sharing approaches to improve local detection. The overview of the information sharing platform is given in Figure~\ref{fig:global}.
 
\subsection{PORTFILER System Overview.}

\portf~\cite{portfiler} is an unsupervised ML threat detection system that uses Zeek network connection logs~\cite{zeek} collected at the border of the enterprise network. This data contains connection metadata information such as timestamp, duration, source and destination IP and port, transport protocol, payload size, number of packets, and connection state. \portf\ monitors a set of configurable ports, including port 23 (Telnet) used by Mirai, port 445 (SMB) used by WannaCry, as well as ports 22 (SSH), 80 (HTTP), and 443 (HTTPS). 


\portf\ learns the distribution of network traffic on each port during the training period, using a set of 35 statistical traffic features aggregated by one-minute time windows (Appendix \ref{app:features}). Ensembles of KDE models are shown to be the most resilient to evasion~\cite{portfiler}.  The models that are part of the ensembles are trained on individual features. During testing, each  model generates an anomaly score. All scores are combined into a final score using: (1)  uniform weights (\textit{Mean Ensemble}); or (2) different weights  (\textit{Weighted Ensemble}).

%






\subsection{Information Sharing for SPM.}

SPM attacks rarely target a single network or organization, due to their ability to quickly propagate across the Internet. Our main hypothesis is that organizations can share threat information after SPM attack detection to help other organizations improve their detection. We propose several scenarios of sharing information between networks. Privacy is a major consideration for threat information sharing and all our methods protect the privacy of security logs, as we only share aggregated information on network traffic, as described below:

\noindent \textbf{Model Sharing.} \neta\ trains the ML model on its network logs, and shares the model directly with \netb. This approach  saves training time at \netb, but differences in traffic patterns between the two sites may render the transferred model less efficient in capturing anomalies at \netb. 


\noindent \textbf{Weight Sharing.} \neta\ and \netb\ train two ensemble models independently on their own  logs.  Assuming that \neta\ is detecting an SPM attack, it computes feature weights based on the detected samples and shares them with \netb.


To compute feature weights, \neta\  applies a Random Forest classifier trained on the malicious samples, as well as legitimate samples to derive feature importance. The feature weights will then be normalized values of feature importance, summing up to 1. \netb\ uses the Weighted Ensemble trained on its own network traffic, with the model weights shared by \neta. Intuitively, the features that are most relevant for the detected attack get assigned higher weights, and will contribute more in the anomaly score generated by the Weighted Ensemble at \netb. We compare the Weighted Ensemble with shared weights to the unsupervised Mean Ensemble with equal weights at \netb.

\noindent \textbf{Weight Adaptation.} Feature distributions on the two networks can be significantly different, and, thus, some feature weights do not transfer effectively. To alleviate this effect, we propose to share information about feature distribution from \neta, in particular the first four moments of distribution of each feature values: mean, variance, skewness and kurtosis. \netb\ compares the distribution of its own features to those of \neta\ and selects only the closest features and their weights in the Weighted Ensemble. The method is outlined below:



\vspace{-5pt}
\begin{enumerate}[noitemsep]
\item Each network computes its moments of distribution: $M_{A}[X]$, $M_{B}[X]$ for each feature vector $X$.
\item \neta\ computes the feature weights.
\item \neta\ shares its feature weights and moments of distribution with \netb.
\item \netb\ computes  Euclidean distance between the moments of the two networks: $||M_{A}[X] - M_{B}[X]||_2$
\end{enumerate}
\vspace{-5pt}
Finally, \netb\ selects top $k$ (i.e., $k=10$) features ordered by distance and carries out detection using
only these top features.



\section{Evaluation}
\label{eval_section}

\subsection{Dataset}
We used Zeek connection logs collected by University of Virginia (UVA) and Virginia Tech (VT). Our experimental setup uses one week of training and one day of testing at each network on July 2020, 9.64 billion events for UVA, and 9.69 billion events for VT. The dataset was anonymized to not reveal personal information about the machines or users on the network\footnote{The IRB office reviewed our data collection process and determined that our research does not qualify as Human Subject Research.}. To evaluate our system, we merge malicious Mirai traces~\cite{stratosphere} at testing time on different ports. We generated an evasive variant of Mirai, 128 times slower than the original, by sampling a fraction of connections to reduce the scanning rate. Ongun et al.~\cite{portfiler} showed that scanning speed may differ among prevalent SPM families.



\subsection{Experiments}

We simulated a scenario where \textsc{Net-A} (UVA) is infected by the original Mirai malware. \textsc{Net-A} runs the unsupervised Mean Ensemble model of \portf\ and detects this attack with accuracy above $0.96$. We assume that \textsc{Net-B} (VT) is infected later either by a similar fast Mirai variant, or by the 128x slower variant. We  evaluated the three proposed information sharing methods from \textsc{Net-A} to \textsc{Net-B}, where \portf\ is also deployed with the same set of features. We start our evaluation with a baseline model where no information sharing is employed and then compare the three sharing methods.



\begin{figure}[th]
	\centering
	\begin{subfigure}[b]{0.49\linewidth}
		\includegraphics[width=\linewidth]{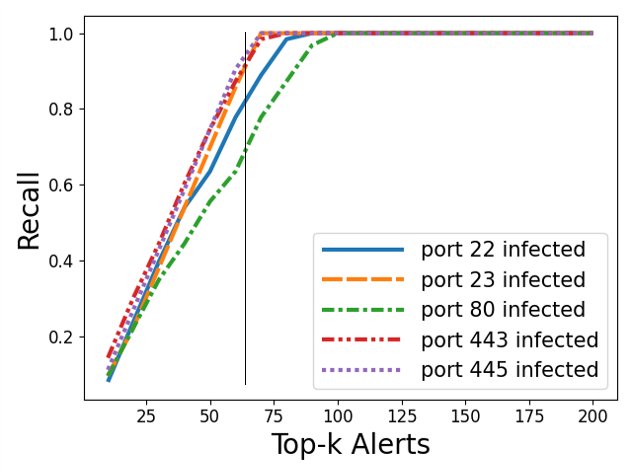}
		\caption{Fast Mirai variant}
		\label{fig:baseline_vt_fast}		
	\end{subfigure}
	\begin{subfigure}[b]{0.49\linewidth}
		\includegraphics[width=\linewidth]{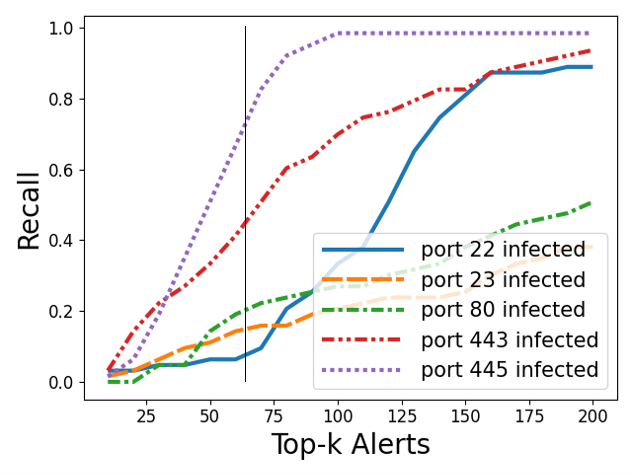}
         \caption{Slow Mirai variant}		
		\label{fig:baseline_vt_slow}
	\end{subfigure}
	\caption{Baseline at \netb, the accuracy is low for slow variant with no shared information. The vertical line represents the number of Mirai samples (63). }
	\label{fig:baseline_vt}
	
	\vspace{-5mm}
\end{figure}

\noindent \textbf{Baseline.} We consider the Mean Ensemble of KDE models trained on \netb, without any shared information.  Figure~\ref{fig:baseline_vt} shows the recall in the top $k$ alerts for the fast and slow Mirai variants. Our test data contains 63 malicious samples (one-minute time intervals) out of 1440 in total. Ideal recall in top $k$ is $k/m$, where $m$ is the number of malicious samples. Therefore, an ideal model would rank all 63 malicious samples on top, without any false positives. This translates into a recall of 1.0 after processing the top 63 alerts. We represent the number of Mirai samples in our plots with a vertical line. The baseline model performs well on the fast Mirai variant (Figure~\ref{fig:baseline_vt_fast}).  However, its performance is much reduced on the slow Mirai variant (Figure~\ref{fig:baseline_vt_slow}) on most ports.


\begin{figure}[th]
	\centering
	\begin{subfigure}[b]{0.49\linewidth}
		\includegraphics[width=\linewidth]{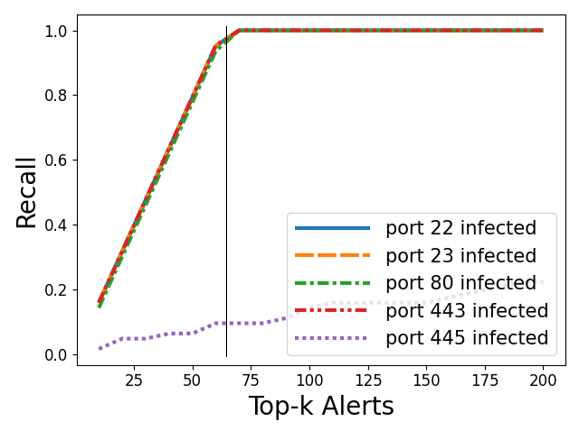}
		\caption{Fast Mirai variant}
		\label{fig:model_sharing_fast}		
	\end{subfigure}
	\begin{subfigure}[b]{0.49\linewidth}
		\includegraphics[width=\linewidth]{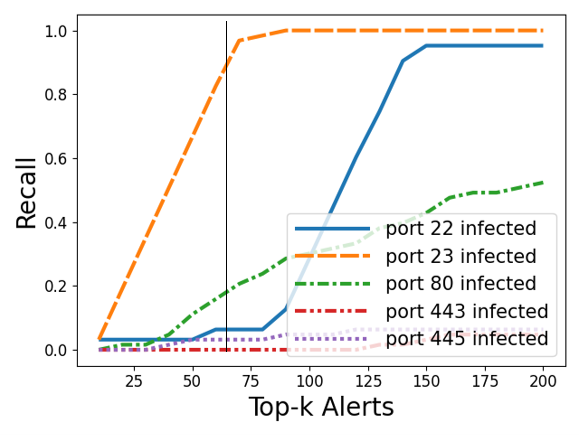}
		\caption{Slow Mirai variant}
		\label{fig:model_sharing_slow}	
	\end{subfigure}
	\caption{Model Sharing. Traffic pattern variations between the two sites where training and testing are done decrease accuracy. The vertical line represents the number of Mirai samples (63). } 
	\label{fig:model_sharing}
		\vspace{-4mm}
\end{figure}

\noindent \textbf{Model Sharing.}
Here, \neta\ shares the entire ML model (Mean Ensemble) and the results at \netb\ are shown in Figure~\ref{fig:model_sharing}. The fast Mirai variant is detected generally well, except on port 445. This is due to different background traffic patterns at the two sites. \neta\ blocks most of the traffic on port 445, while \netb\ still allows internal  traffic on this port. This leads to an important observation: on some ports, the model is not directly transferable between sites. Prior analysis of the traffic patterns is necessary to establish whether the model can be shared. For the slower Mirai variant, we generally see a performance degradation when the model is shared compared to the baseline. At slower malware propagation speeds, the detector is more sensitive to traffic variations between the networks.


\begin{figure}[th]
	\centering
	\begin{subfigure}[b]{0.49\linewidth}
		\includegraphics[width=\linewidth]{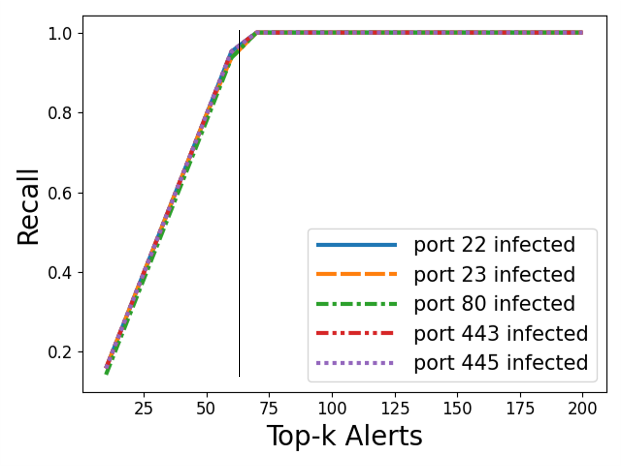}
		\caption{Fast Mirai variant}
		\label{fig:weight_sharing_fast}	
	\end{subfigure}
	\begin{subfigure}[b]{0.49\linewidth}
		\includegraphics[width=\linewidth]{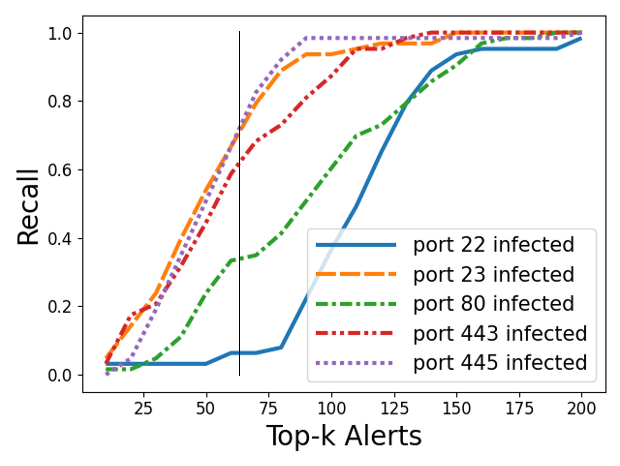}
		\caption{Slow Mirai variant}
		\label{fig:weight_sharing_slow}	
	\end{subfigure}
	\caption{Weight Sharing. Detection is improved by giving higher weights to features that are more relevant to a particular attack. Mirai samples = 63 (vertical line)}
	\label{fig:weight_sharing}
	\vspace{-5mm}
\end{figure}

\noindent \textbf{Weight Sharing.} Each site trains an ensemble model on their own data. \neta\ uses the Mean Ensemble detector to rank and label malicious time windows and derive feature weights. These weights are shared with \netb, which uses its own trained model, along with these shared weights to detect anomalies in the test data. Figure~\ref{fig:weight_sharing} shows the detector's performance on \netb\ test data. This method is able to correctly identify malicious time windows better than either the baseline or the model sharing scenario.  It obtains almost perfect detection on the fast Mirai variant, and generally better results on the slower Mirai variant. However, on ports 22 and 80 the results can be further improved.

\begin{figure}[th]
	\centering
	\includegraphics[width=0.49\linewidth]{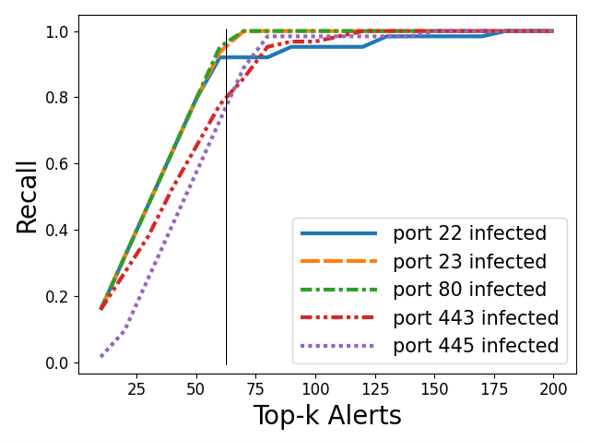}

	\caption{Weight Adaptation (slow variant). This method performs best, by adapting the shared weights based on the statistical distances between features at the two sites. Mirai samples = 63 (vertical line).}
	\label{fig:weight_adaptation}
	\vspace{-5mm}
	
\end{figure}

\noindent \textbf{Weight Adaptation.}
We take the weight sharing approach one step further by adjusting the weights based on the statistical distance between feature distributions at the two sites and using only the 10 closest features. The fast Mirai attack is detected with maximum recall and precision and thus the graph has been omitted. Figure~\ref{fig:weight_adaptation} shows the system's ability to detect malicious time windows at \netb\ on the slow Mirai attack.  The recall metric is improved across all ports, illustrating the benefit of an adaptive approach for weight sharing. Precision and false positive comparison of sharing methods are presented in Table~\ref{tab:prec_top_60} and~\ref{tab:fp_top_60} of Appendix~\ref{app:comparison}.

\section{Related Work}

Vasilomanolakis et al.~\cite{vasilomanolakis2015taxonomy} provide a  taxonomy of collaborative intrusion detection systems. Earlier work on distributed intrusion detection leverages matching attack graph signatures to detect coordinated attacks~\cite{yang2000cards, ning2001abstraction}. Alert correlation has been proposed to reduce the number of false positives by sharing alerts centrally to confirm malicious behavior~\cite{cuppens2002alert, boggs2011cross}.
Hu et al.~\cite{hu2013online} uses boosting classifiers for distributed threat detection. Parameterized local models are collected to build a global model improving the detection rates.
Nguyen et al.~\cite{nguyen2019diot} study the first federated-learning-based anomaly detection system for IoT networks.



Threat intelligence sharing has been studied to consider different challenges such as efficient coordination, addressing legal regulations, and standardization efforts~\cite{skopik2016problem, kampanakis2014security, dandurand2013towards}.


\section{Discussion and Conclusion}

We proposed three information sharing methods across two networks and showed how the shared information can be used in an ML model to significantly improve its ability to detect evasive self-propagating malware. We showed that network traffic distribution differs significantly per port across organizations and direct model transfer is not effective. While weight sharing works better than model sharing, the weight adaptation method, which selects feature weights according to the closest features in the two networks, performs best at detecting the evasive variant.



We highlight the need for global defensive models that can benefit from information sharing across defenders to better protect against coordinated adversaries. We discuss several challenges and open problems to realize this vision:



\noindent \textbf{Generalization and Evasion.} Although we show the effectiveness of our approach against SPM attacks, generalization to other attacks remains an open question. Can the  ensemble model  be used with different weights to adapt to other attacks? We have shown resilience to evasion against an adversary slowing down the propagation rate, but how resilient are these methods against more advanced evasive strategies? 




\noindent \textbf{Threat intelligence sharing.} Platforms such as MISP~\cite{misp} facilitate threat sharing of IPs, domains, and malware file hashes. These indicators help detect attacks with known behavior, but how well do they work for new attack campaigns or even attack variants? It is relatively easy for attackers to rotate their domain and IP infrastructures to evade static indicators. Determining a set of malware invariants  to detect evolving attacks is one of the challenges in this space.


\noindent \textbf{Multiple Parties.} We experimented with information sharing between two networks, but can information sharing across multiple defenders provide an opportunity to develop more resilient ML detectors? Possible architectures include centralized threat sharing platforms (similar to MIPS), and peer-to-peer models. The P2P model offers a more flexible trust model, while the centralized model provides more coordination and a global perspective on the threat landscape. An important open question is how to determine which information is trustworthy and how to detect potential attackers that manipulate the shared information.

\noindent \textbf{Privacy considerations.} When designing threat information sharing methods, privacy of the security logs needs to be maintained. Federated learning provides decentralized learning methods to aggregate local model parameters trained by individual clients into a global ML model. Its applicability to ML models for threat detection is an interesting topic of future work.





\section*{Acknowledgements}

This research was sponsored by the contract
number W911NF-18-C0019 with the U.S. Army Contracting Command - Aberdeen Proving Ground (ACC-APG) and the Defense Advanced Research Projects Agency (DARPA), and 
by the U.S. Army Combat Capabilities Development Command
Army Research Laboratory under Cooperative Agreement Number
W911NF-13-2-0045 (ARL Cyber Security CRA). The views and conclusions contained in this document are those of the authors and should not be interpreted as representing the official policies, either expressed or implied, of the ACC-APG, DARPA, Combat Capabilities Development Command Army Research Laboratory or the U.S. Government. The U.S. Government is authorized to reproduce and distribute reprints for Government purposes notwithstanding any copyright notation here on. This project
was also funded by NSF under grant CNS-171763. 

\bibliographystyle{abbrv}
\bibliography{evasion,malware}

\appendix

\section{Additional Experiments.}
We also conducted several experiments intended to further understand why the slow Mirai variant exhibits better recall on some ports compared to others in Figure~\ref{fig:weight_sharing_slow}.

To this end, we aggregated the feature distances to produce a single distance metric per port. We experimented with two aggregation methods: a simple average and a weighted average where the weights are the feature importance coefficients. We analyzed whether weight sharing performs better on ports with shortest aggregated distance between the two networks.

Before aggregation, distances per feature were computed with the following methods:
\begin{itemize}[noitemsep]
\item Euclidean distance over the first four moments of distribution as described in the methodology section.

\item Euclidean distance over scale-adjusted moments of distribution. Given that the orders of magnitude of mean, variance, skewness and kurtosis are $E[X]$, $E[X^2]$, $E[X^3$] and $E[X^4]$, respectively, we compute 
$E[X], E[X^2]^{1/2}, E[X^3]^{1/3},  E[X^4]^{1/4}$ to bring them to the same scale.

\item For completeness, we also experimented with Earth mover's distance (Wasserstein metric) between feature distributions on the two networks, instead of using the moments of distribution. This method is not intended for practical uses, due to the difficulty of transferring entire feature distributions from one network to the other. 

\end{itemize}

While these experiments helped us better understand the data, they were not conclusive in explaining the difference in performance on the five ports in Figure~\ref{fig:weight_sharing_slow}. Since traffic patterns and feature distributions are quite complex, a single aggregated distance metric may not capture the necessary information. Determining on which ports the weights transfer better is more subtle and requires further research.

\section{Comparison of Sharing Methods.}
\label{app:comparison}
We compare the three methods proposed using various performance metrics. We have previously discussed the recall metric in Section~\ref{eval_section}. In this section, we look at precision and false positives for each of the five ports on the slow Mirai variant.

Table~\ref{tab:prec_top_60} illustrates performance in terms of precision, while Table~\ref{tab:fp_top_60} illustrates performance in terms of false positives in top-60 alerts.

As these tables show, the Weight Sharing method generally improves on the Model Sharing method. The Weight Adaptation method delivers best precision with fewest false positives across the board. 

\begin{table}[t]
		\centering
		\scalebox{0.8}{
			\begin{tabular}{c|c|c|c|c|c|}
				\cline{2-6}
				\multicolumn{1}{l|}{}                        & \multicolumn{5}{c|}{\textbf{Infected Port}} \\  \cline{2-6}
				\multicolumn{1}{l|}{} & \textbf{80}     & \textbf{443}    & \textbf{22}           & \textbf{23}        & \textbf{445}    \\ \hline
				\multicolumn{1}{|c|}{\textbf{Model Sharing}}            &            0.16  &  0              & 0.06                  & 0.86     &  0.03           \\ \hline
				\multicolumn{1}{|c|}{\textbf{Weight Sharing}}           & 0.35    &  0.61              &  0.06                 & 0.70     &  0.70            \\ \hline
				\multicolumn{1}{|c|}{\textbf{Weight Adaptation}}            & 1.0            & 0.81          & 0.96                  & 0.98  & 0.76                 \\ \hline
		\end{tabular}}
		\captionsetup{margin=1cm}
		\caption{Performance in terms of Precision in the top-60 ranked alerts across the ports, for the slow Mirai variant at VT.}
		\label{tab:prec_top_60}
\end{table}
\begin{table}[t]
		\centering
		
		\scalebox{0.8}{
			\begin{tabular}{c|c|c|c|c|c|}
				\cline{2-6}
				\multicolumn{1}{l|}{}                        & \multicolumn{5}{c|}{\textbf{Infected Port}} \\ \cline{2-6}
				\multicolumn{1}{l|}{} & \textbf{80}     & \textbf{443}    & \textbf{22}           & \textbf{23}        & \textbf{445}    \\ \hline
				\multicolumn{1}{|c|}{\textbf{Model Sharing}}            & 50             & 60               &   56               &  8   &   58          \\ \hline
				\multicolumn{1}{|c|}{\textbf{Weight Sharing}}           & 39             &  23              &  56                 & 18     &   18           \\ \hline
				\multicolumn{1}{|c|}{\textbf{Weight Adaptation}}            & 0            & 11          & 2                  & 1  & 14                 \\ \hline
		\end{tabular}}
		\captionsetup{margin=1cm}
		\caption{Performance in terms of False Positives in the top-60 ranked alerts across the ports, for the slow Mirai variant at VT.}
		\label{tab:fp_top_60}
\end{table}

\section{Traffic Features.}
\label{app:features}

We present a high-level overview of features used in \portf\ in four categories. We extract these features for each port separately. We define these features for each 1-minute time window. The complete list of features is described in \cite{portfiler}.

 \begin{enumerate}[leftmargin=0.3cm]
	\item {\bf Traffic statistics features:} We extract the number of distinct internal and external IPs communicating on that port, number of connections, and number of new distinct external IPs.
	\item {\bf Duration features}: We extract max, min, variance and mean of duration values.
	\item {\bf Bytes and packets features}: We extract max, variance, and mean of sent and received bytes and packets values. We also define the number of connections with no bytes received as a separate feature. 
	\item {\bf Connection state features:} We extract the number of connections for each Zeek connection state string (S0, S1, OTH, etc.). We also define the number of failed connections as a separate feature. 
\end{enumerate}

\end{document}